\documentclass[%
 reprint,
superscriptaddress,
 amsmath,amssymb,
 aps,
prl,
]{revtex4-2}

\usepackage{graphicx}
\usepackage{float}
\usepackage{dcolumn}
\usepackage{bm}
\usepackage{hyperref}
\hypersetup{colorlinks,linkcolor={blue},citecolor={blue},urlcolor={blue}}  
\usepackage{xcolor} 
\usepackage{fixltx2e}

\begin{document}


\title{
Element selective ultrafast magnetization dynamics  of hybrid Stoner-Heisenberg magnets}
 

\author{Mohamed F. Elhanoty}
\email{mohamed.elhanoty@physics.uu.se}
 \affiliation{Division of Materials Theory, Department of Physics and Astronomy, Uppsala University, Box-516, SE 75120, Sweden}
 

\author{Olle Eriksson}
\affiliation{Division of Materials Theory, Department of Physics and Astronomy, Uppsala University, Box-516, SE 75120, Sweden}
\affiliation{School of Science and Technology, Örebro University, SE-701 82, Örebro, Sweden}

\author{Ronny Knut}
\affiliation{Division of Materials Theory, Department of Physics and Astronomy, Uppsala University, Box-516, SE 75120, Sweden}

\author{Olof Karis}
\affiliation{Division of Materials Theory, Department of Physics and Astronomy, Uppsala University, Box-516, SE 75120, Sweden}

\author{Oscar Gr\aa n\"as}
\email{Corresponding Author: oscar.granas@physics.uu.se}
\affiliation{Division of Materials Theory, Department of Physics and Astronomy, Uppsala University, Box-516, SE 75120, Sweden}

\date{\today}

\begin{abstract}
Stoner and Heisenberg excitations in magnetic materials are inherently different. The former involves an effective reduction of the exchange splitting, whereas the latter comprises excitation of spin-waves. In this work, we test the impact of these two  excitations in the hybrid Stoner-Heisenberg system FePd. 
We present a microscopic picture of ultrafast demagnetization dynamics in this alloy, which represents both components of strong local exchange splitting in Fe, and induced polarization in Pd. 
We identify spin-orbit coupling  and  optical  inter-site  spin  transfer as the two dominant factors for demagnetization at ultrashort timescales. By  tuning  the  external  laser  pulse, the extrinsic inter-site spin transfer can be manipulated for site selective demagnetization on 
femtosecond time scales providing the fastest way for optical and selective control of the magnetization dynamics in alloys. Remarkably, the drastic difference in origin of the magnetic moment of the Fe and Pd species is not deciding the initial magnetization dynamics in this alloy. 

\end{abstract} 
\maketitle



Laser induced ultra-fast magnetization dynamics (LIUMD) has been demonstrated to be an efficient way to manipulate the spin and orbital moments of magnetic elements in the femtosecond regime \cite{kimel2007femtosecond,beaurepaire1996ultrafast}.  Manipulation of the demagnetization amplitude and the corresponding time scale in an element specific way, for multi-component systems, is of fundamental interest to further engineer the LIUMD technique\cite{hofherr2020ultrafast, stanciu2007all}. Although the experimental realizations are well established, theory is lagging behind. Several theoretical  attempts  to explain the interplay between various degrees of freedom in the demagnetization process 
have been suggested\cite{koopmans_microscopic_2005, koopmans_unifying_2005, dewhurst_where_2020,elliott_microscopic_2020, hohlfeld1997nonequilibrium,aeschlimann1997ultrafast,carpene2008dynamics,dewhurst_laser-induced_2018, tengdin2020direct,chimata2015all,malik2021ultrafast,battiato_superdiffusive_2010}, all however showing deficiencies in explaining the full experimental results.

Most systems that have been investigated so far, both by experiments and theory, are known to be well represented by a Heisenberg Hamiltonian, possibly with a correction from anisotropic exchange (e.g. Dzyaloshinskii-Moriya interaction). At longer time scales, e.g. those that are relevant for magnon excitations, these systems can be interpreted from a low energy spin-Hamiltonian, and this has motivated an analysis of the ultrafast experiments reported in Refs.\cite{beaurepaire1996ultrafast} by means of atomistic spin-dynamics \cite{chimata2015all, eriksson2017atomistic, evans2015quantitative}.
However, magnetic systems that known to have Stoner excitations, are much less studied. A key to understanding the LIUMD process in general, is to compare the excitation processes in Heisenberg and Stoner systems on equal footing. 

Elemental Pd is on the verge of being ferromagnetic due to its high density of states (DOS) at the Fermi level \cite{hoare_low-temperature_1957} and a Stoner product just below one. Alloys between Pd and magnetic 3d elements such as Fe or Co represent unconventional magnets, with an experimentally observed "giant" magnetic moment due to the induced polarization  the dilute limit of Fe or Co impurities in Pd \cite{wolff_ferromagnetism_1961,low_distribution_1966}.
To be specific, it was found experimentally that Co and Fe induce magnetic moments on large numbers of nearby Pd atoms, with an effective net moment of 10 $\mu_B$ to 12 $\mu_B$ per 3d impurity atom. This giant moment stems from an enhanced local moment of the Co and Fe atoms, compared to their elemental values\cite{low_distribution_1966}, in combination with a cloud of induced moments of the Pd atoms that are in the vicinity of the impurity host. The induced Pd moment is for all systems less than 0.2 $\mu_B$/atom\cite{mohn_supercell_1993}, but since the induced moment extends many atomic shells away from the 3d impurity, each impurity 3d atom is seen to correspond to a total moment of 10 - 12 $\mu_B$. 

The giant moment in 3d-Pd alloys was also demonstrated by ab initio calculations\cite{mohn_supercell_1993}. In this work it was also shown that the magnetic excitations of the Pd moment in these systems can be described by a Landau expansion of the free energy in terms of the order parameter. One can therefore conclude that alloys between magnetic 3d elements (Cr, Mn, Fe, Co and Ni) and Pd, especially in the dilute limit, are ideal to investigate the phenomenon of LIUMD of a multicomponent system with building blocks that have drastically different magnetic properties and mechanisms of excitations. To be specific, the local moment of Fe in a Pd matrix, has excitations that are expected to be described by the Heisenberg Hamiltonian, while the Pd  expected to undergo Stoner excitations. This drastic difference is expected also for higher concentrations of 3d elements, such as in ordered compounds like FePd. In fact, one experimental investigation of light induced modification of the magnetism has been published for FePd\cite{iihama2015ultrafast}, but  element specific analysis of the magnetism was not made in this work, nor was the expected difference in excitation properties analyzed. 

Time dependent density functional theory (TDDFT) is an ab-initio theory that can be used to study LIUMD, where only the composition and geometry of the material are needed as input, in combination with the experimental parameters for the laser pulse\cite{runge1984density}. 
The method has shown great success in addressing the fundamental questions of the early parts of the magnetization dynamics in LIUMD experiments; very good agreement was obtained with the MOKE and XMCD experiments for several systems \cite{dewhurst_laser-induced_2018,dewhurst_element_2020,dewhurst_laser-induced_2018,elliott_ultrafast_2016,yao_distinct_2020}. Both the relevant macroscopic experimental observables, such as magnetization, and microscopic details, such as orbital and spin resolved particle occupations, are extracted from TDDFT calculations, for instance as implemented in the ELK code \cite{elk}.


Previously, multi-component magnetic alloys, such as Heusler compounds, were studied using TDDFT. In these works, spin flips mediated by spin-orbit coupling (SOC), and their importance for the optical inter-site spin transfer (OISTR) effect, have been suggested to be the main source of demagnetization, at least for the early parts of the magnetisation dynamics \cite{stamenova_role_2016,dewhurst_laser-induced_2018}. While the time scale at which the SOC starts to cause spin flips is material and structure dependent, the time scale at which OISTR can produce electronic excitation is also dependent on the optical properties of the laser pulse (e.g full width at half maximum-FWHM, the light frequency and its intensity) \cite{dewhurst_laser-induced_2018}.

The discussion above motivates an investigation of the Fe-Pd alloy using TDDFT. We have chosen FePd$_3$ as a representative, because it contains the essential features of combining Stoner and Heisenberg excitations in the same compound, while providing a computationally manageable system.
In making this choice one also mixes elements with expected, different site projected electronic structures, as well as different spin-orbit strengths. As we shall see below, this offers key parameters to understand the LIUMD effect in this system. With this in mind, the aim of this communication is threefold. First, the elemental contribution to the ultrafast demagnetization of LIUMD of Fe-Pd systems is resolved and a theoretical explanation of its microscopic origin is provided using TDDFT. 
Second, the influence of SOC, combined with differences in on-site projected electronic structures, are analyzed for the demagnetization FePd$_3$. Third, by means of optical control of the laser pulse, a general route to  selectively manipulate the onset of the  demagnetization of individual  elements in a multicomponent system is highlighted. 

TDDFT is an extension of the ground state (GS) DFT formalism, through the one to one mapping between the external potential $v_{ext}(\mathbf{r},t)$ and the electronic density $n$($\mathbf{r}$,t) that only depends on spatial coordinates in the GS, while on both spatial coordinates and time in  TDDFT. This unique one to one correspondence allows for a fully interacting system to be mapped into an equivalent non-interacting one with a time dependent Kohn-Sham (KS) effective potential $v_s$($\mathbf{r}$,t), that produces the same density as the fully interacting system, during every time propagation step. Notably, $v_s(\mathbf{r},t)$ is a sum of three terms $v_s(\mathbf{r},t)=v_{ext}(\mathbf{r},t)+v_{H}(\mathbf{r},t)+v_{xc}(\mathbf{r},t)$, where $v_{ext}(\mathbf{r},t)$ is the external potential, $v_{H}(\mathbf{r},t)$ is the Hartree potential, and  $v_{xc}(\mathbf{r},t)$ is the exchange-correlation (XC) potential. Within TDDFT, the effect of an external laser pulse is treated within the dipole approximation. This approximation is valid for radiation  with long wavelength compared to the lattice constant. For all pump-probe experiments done currently, this is an excellent approximation. This allows to consider only a time dependent vector potential $\mathbf{A}$\textsubscript{ext}(t) that modifies the kinetic energy  (first term) of the Hamiltonian in Eq. \ref{eq:TDSERG}
\begin{equation}
\begin{split}
    \bigg[\frac{1}{2}{\big (} -i\nabla + \frac{1}{c} \mathbf{A}_{ext}(t){\big )}^2+v_s(\mathbf{r},t)
    +\frac{1}{2c}\sigma \cdot \mathbf{B}_s(\mathbf{r},t)+ \\
    \frac{1}{4c^2} \sigma \cdot {\big (}\nabla v_s(\mathbf{r},t)\times -i\nabla{\big )} 
    \bigg] \psi_i(\mathbf{r},t)=\frac{\partial \psi_i(\mathbf{r},t)}{\partial t},
\end{split}
  \label{eq:TDSERG}
\end{equation}
where c is the speed of light, $\sigma$ is the Pauli matrix, and $\mathbf{B}_s(\mathbf{r},t)$ is the effective KS magnetic field ($\mathbf{B}_s(\mathbf{r},t)=\mathbf{B}_{ext}(t)+\mathbf{B}_{xc}(\mathbf{r},t)$, where $\mathbf{B}_{ext}(t)$ is the magnetic field of the external laser pulse and $\mathbf{B}_{xc}(\mathbf{r},t)$ is the XC induced exchange splitting, expressed in form of a field. The last term of Eq. \ref{eq:TDSERG} is the SOC term, and $\psi_i(\mathbf{r},t)$ is two component Pauli spinor. 
The atomic units are adopted in all the equations through this paper with $\hbar=e=$m=1. 

The  Hamiltonian of Eq. \ref{eq:TDSERG} is diagonalized in a Linearized Augmented Plane Wave (LAPW) basis set with two variational steps. In the first step, the Hamiltonian containing only the scalar potential is diagonalized in the LAPW basis. In the second variational step, a sufficient number of the resulting scalar states are  used as a basis to construct
the Hamiltonian with spinor degrees of
freedom. The core electrons are treated within the radial Dirac equation, and the valence electrons are treated within the relativistic Hamiltonian in the presence of the SOC.

The occupation of the KS transient state $\eta(\epsilon,t)$ discussed below 
is calculated by first projecting the TDKS state, $\psi(r,t)$, on the time independent, ground state, $\phi(r)$, via
\begin{equation}
    P_{ij}^\mathbf{k}(t)=\int d^3r \phi^\star_{i\mathbf{k}}(r)\psi_{j\mathbf{k}}(r,t)
    \label{eq:projection}
\end{equation}
Summing the square of the the projection $P_{ij}^\mathbf{k}(t)$ over all time dependent KS states, weighted by the occupation number, $n_{jk}$, gives the TD occupation projected on the ground state 
\begin{equation}
    w_{i\mathbf{k}}(t)=\sum_j n_{jk}|P_{ij}^\mathbf{k}(t)|^2.
    \label{eq:occ_projection}
\end{equation}
Finally, the TD projected DOS, $\eta(\epsilon,t)$, is evaluated according to 
\begin{equation}
    \eta(\epsilon,t)=\sum_i^\infty  \int_{BZ}\delta(\epsilon-\epsilon_{i\mathbf{k}}) w_{i\mathbf{k}}(t),
    \label{eq:td_pdos}
\end{equation}
where $\epsilon_{i\mathbf{k}}$ is the i$^{th}$ Kohn-Sham energy eigenvalue. 

\begin{figure}[H]
    \centering
    \includegraphics[width=\linewidth]{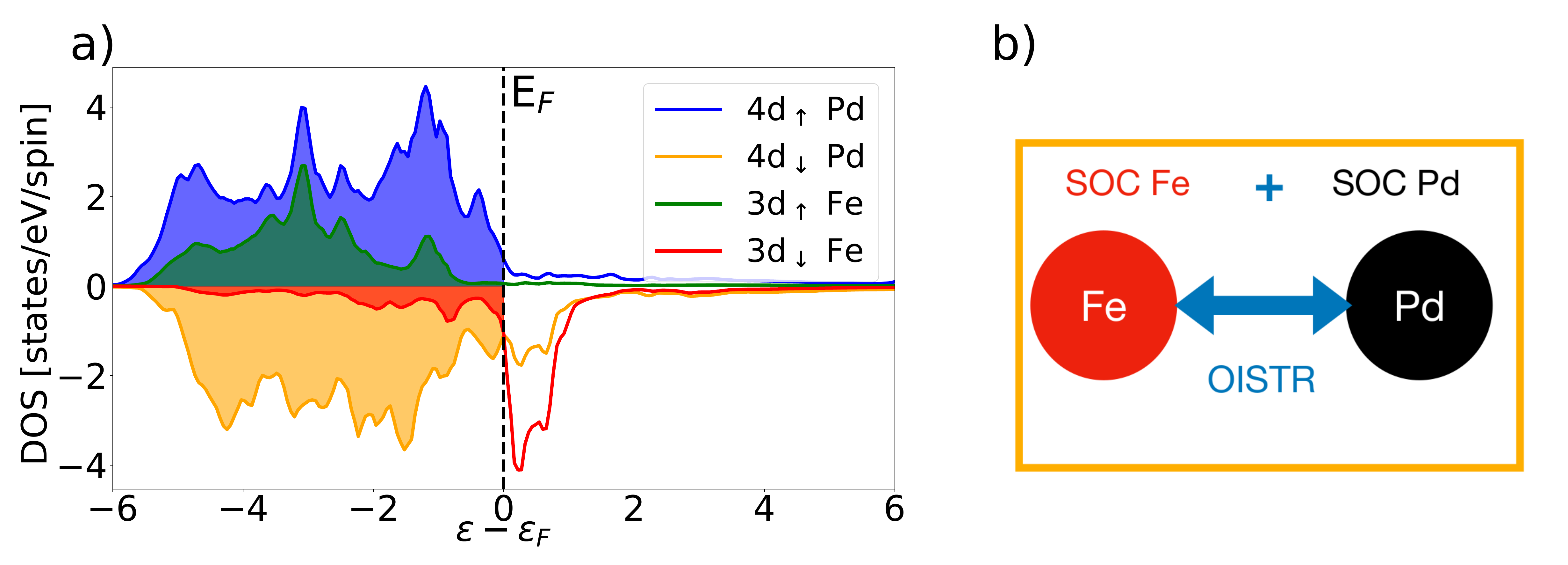}
    \caption{(\textbf{a}) GS DOS (full lines) and occupations (shaded colors) of the 4d and 3d levels for  the three Pd atoms and Fe respectively. (\textbf{b}) Schematic representations for the available source of demagnetization in FePd alloy.}
    \label{fig:gs_d_orbital_dos}
\end{figure}

The structure is cubic, with space group is $Pm{\bar 3}m$. The calculations of this system were performed in a fully ab-initio and noncolinear fashion, as implemented in  the ELK code, with 8 $\times$ 8 $\times$ 8 k points in the Brillouin-zone. The calculations were performed in three steps. First, the geometry of the FePd$_3$ system was optimized. Second, the ground state (GS) of the optimized  structure was determined and partial DOS (pDOS) of the d-states (shown in Fig. \ref{fig:gs_d_orbital_dos}\textcolor{blue}{a}) and their occupations were evaluated. Third, the optimized GS of the structure was evolved in time, using Eq. \ref{eq:TDSERG}, with a time step of 2.4 attoseconds and a time propagation scheme as described in reference \cite{dewhurst_efficient_2016}. A laser pulse with wavelength of 800 nm and fluence of 1 mJ/cm$^2$ and full width at half maximum (FWHM) of 40 fs was allowed to interact with the electronic subsystem, and the response to this external field was followed for over 140 fs. 

The local spin density approximation (LSDA) and its time-dependent, adiabatic version (ALSDA), for exchange and correlation, was used in the ground and transient states, respectively. Secondary scattering processes such as the nuclei and radioactive contribution to dynamics were ignored. These effects are assumed to not influence the demagnetization at the early times of the process, as the dynamics is mainly driven by the electronic degrees of freedom, coupled to the laser pulse \cite{fann1992electron,sun1994femtosecond,suarez1995dynamics}. 



\begin{figure}
    \centering
    \includegraphics[width=\linewidth]{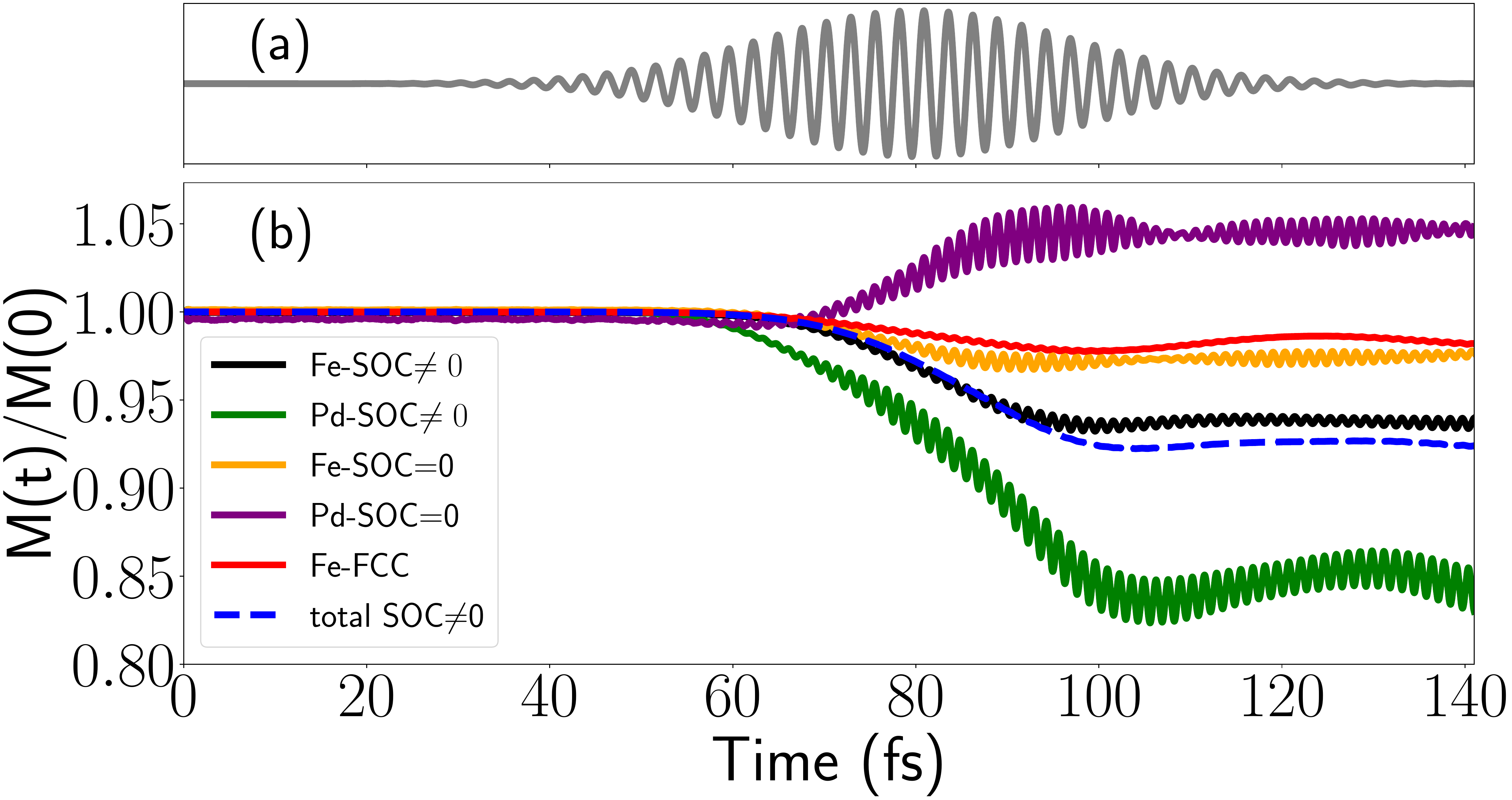}
    \caption{(\textbf{a}) Pulse shape of the laser with FWHM=40 fs.(\textbf{b}) Comparison between the Fe and Pd moments in Fe3Pd alloy for SOC=0 and SOC $\ne$0 and for Pure FCC Fe (see the text).}
    \label{fig:fepd_all_mom_compar}
\end{figure}

The calculated total GS magnetic moment of the system is 4.18 $\mu_B$, distributed as 3.23 $\mu_B$ on Fe and 0.32 $\mu_B$ on every Pd atom. This result is consistent with expectations from all previous studies of the Fe-Pd system. The driven magnetization dynamics is shown in Figure \ref{fig:fepd_all_mom_compar}\textcolor{blue}{b}, where we compare the relative demagnetization of Fe and Pd of FePd$_3$ to that of elemental Fe in the fcc structure, using the same laser pulse parameters (the pulse is illustrated in Figure \ref{fig:fepd_all_mom_compar}\textcolor{blue}{a}). Selecting fcc Fe as a comparison material, instead of bcc Fe, is motivated from geometrical reasons, since replacing all Pd atoms with Fe in FePd$_3$ results in the fcc structure (note that for fcc Fe we obtain ground state magnetic moments of 2.77 $\mu_B$ per atom). The total driven demagnetization in Figure \ref{fig:fepd_all_mom_compar}\textcolor{blue}{b} is in very good agreement with the one from the pump probe experiment  reported  in Ref \cite{iihama2015ultrafast}, where $\approx 6\%$ of the total moment is lost.  

Figure \ref{fig:fepd_all_mom_compar}\textcolor{blue}{b} shows
that Fe and Pd atoms in FePd$_3$ demagnetize very differently. The relative demagnetisation of Pd is much stronger than that of Fe. 
Figure \ref{fig:fepd_all_mom_compar}\textcolor{blue}{b} also shows that the relative change of the Fe moment in FePd$_3$ is significantly larger, compared with the loss of the Fe moment in elemental Fe. To further shine light on the mechanisms that are responsible for the  LIUMD effect of FePd$_3$, we performed calculations in which SOC was neglected. Interestingly, as Fig. \ref{fig:fepd_all_mom_compar}\textcolor{blue}{b} shows, the light induced moment of Pd is then found to increase, while the Fe moment decreases, seemingly in a similar way as that of fcc Fe.

Remarkably, the amount of Fe demagnetization in calculations where SOC is included (black graph) is significantly greater than for calculations when SOC=0 (yellow graph). Notably, the role of SOC on the demagnetization of Pd is more profound than for Fe. Pd is seen to loose about 15$\%$ of its initial moment when SOC is included while its moment increases if SOC is neglected. The magnetic moments in FePd$_3$ hence demagnetize through two channels: the OISTR mechanism, which is most clearly seen from calculations without SOC, and from spin flip transitions, visible from calculations with finite SOC. 
These two contributions are schematically shown in Fig. \ref{fig:gs_d_orbital_dos}\textcolor{blue}{b}. 

To analyze the results in Fig.\ref{fig:fepd_all_mom_compar}\textcolor{blue}{b}, we inspect the pDOS curves, shown in Fig.\ref{fig:gs_d_orbital_dos}\textcolor{blue}{a}. It should be noted that the  unoccupied states of the minority spin channel (spin down, $\downarrow$) have special significance, since they have been discussed to be relevant for the  OISTR mechanism. Figure \ref{fig:gs_d_orbital_dos}\textcolor{blue}{a} shows both the occupied and the unoccupied states for the 4d and 3d levels of Pd and Fe atom. The $3d_\downarrow$ states of Fe can be seen to have a significant amount of unoccupied states. These are states available for the OISTR mechanism, that can be thought of as  optical excitations (in the dipole approximation). Significantly fewer states are available for the unoccupied $4d_\downarrow$ states of Pd. A pure OISTR process is best illustrated by the data in Fig.\ref{fig:fepd_all_mom_compar} where spin-orbit coupling is neglected, since spin-flip transitions are absent on this level of theory. Optical excitations then happen within each spin channel separately. As Fig. \ref{fig:gs_d_orbital_dos}\textcolor{blue}{a} shows, optical excitations are more efficient in the spin down channel, since the number of unoccupied states available in this process is larger. Fig. \ref{fig:gs_d_orbital_dos}\textcolor{blue}{a} also shows that the occupied spin down states have more pure Fe character while the unoccupied states have mixed, hybridized Fe and Pd character. The optical excitations can therefore be seen as transitions 
from the occupied minority spins of Pd to the unoccupied minority spin channel of Fe. This leads to a decrease of the Fe moment and an increase of the Pd moment, as shown in Fig.\ref{fig:fepd_all_mom_compar}\textcolor{blue}{b}.
Note that this process, as discussed also in Ref.\cite{tengdin2020direct}, has to occur for hybridized states, since otherwise dipole transitions would not be allowed.

\begin{figure}
    \centering
    \includegraphics[width=\linewidth]{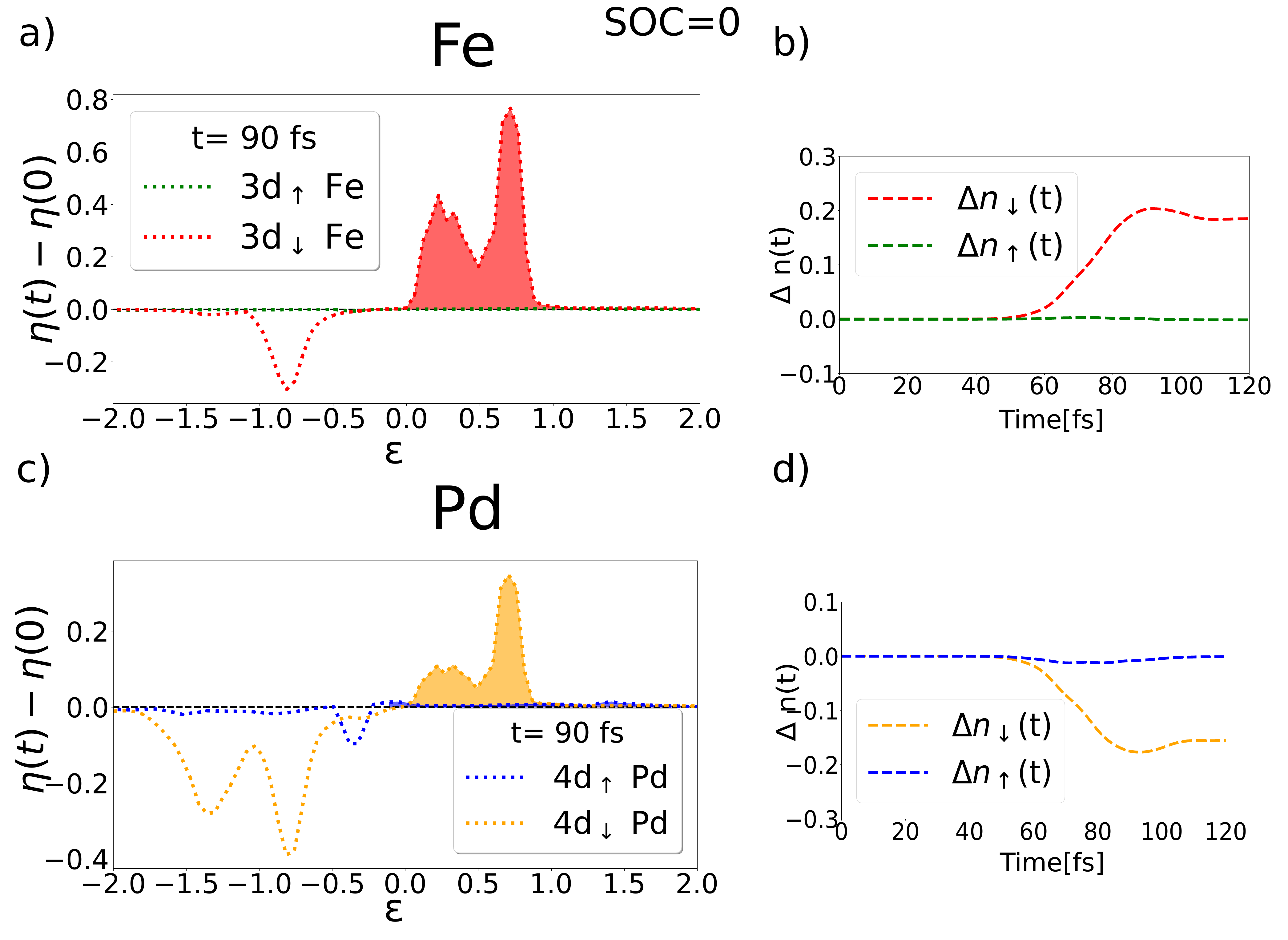}
    \caption{Calculated results with SOC ignored. (\textbf{a,c}) the change in the occupation of the 3d up and down spin channels to the GS ($\eta(0)$) at t=90 fs for Fe and Pd. (\textbf{b,d}) the change in the integration of the majority($\Delta n_\uparrow$(t)=$\int d\epsilon (\eta_\uparrow(\epsilon,t)-\eta_\uparrow(\epsilon,0))$) and minority($\Delta n_\downarrow$(t)) spins occupations over the entire energy range in time for Fe and Pd.}
    \label{fig:fepd_pdos_nosoc}
\end{figure}

\begin{figure}
    \centering
    \includegraphics[width=\linewidth]{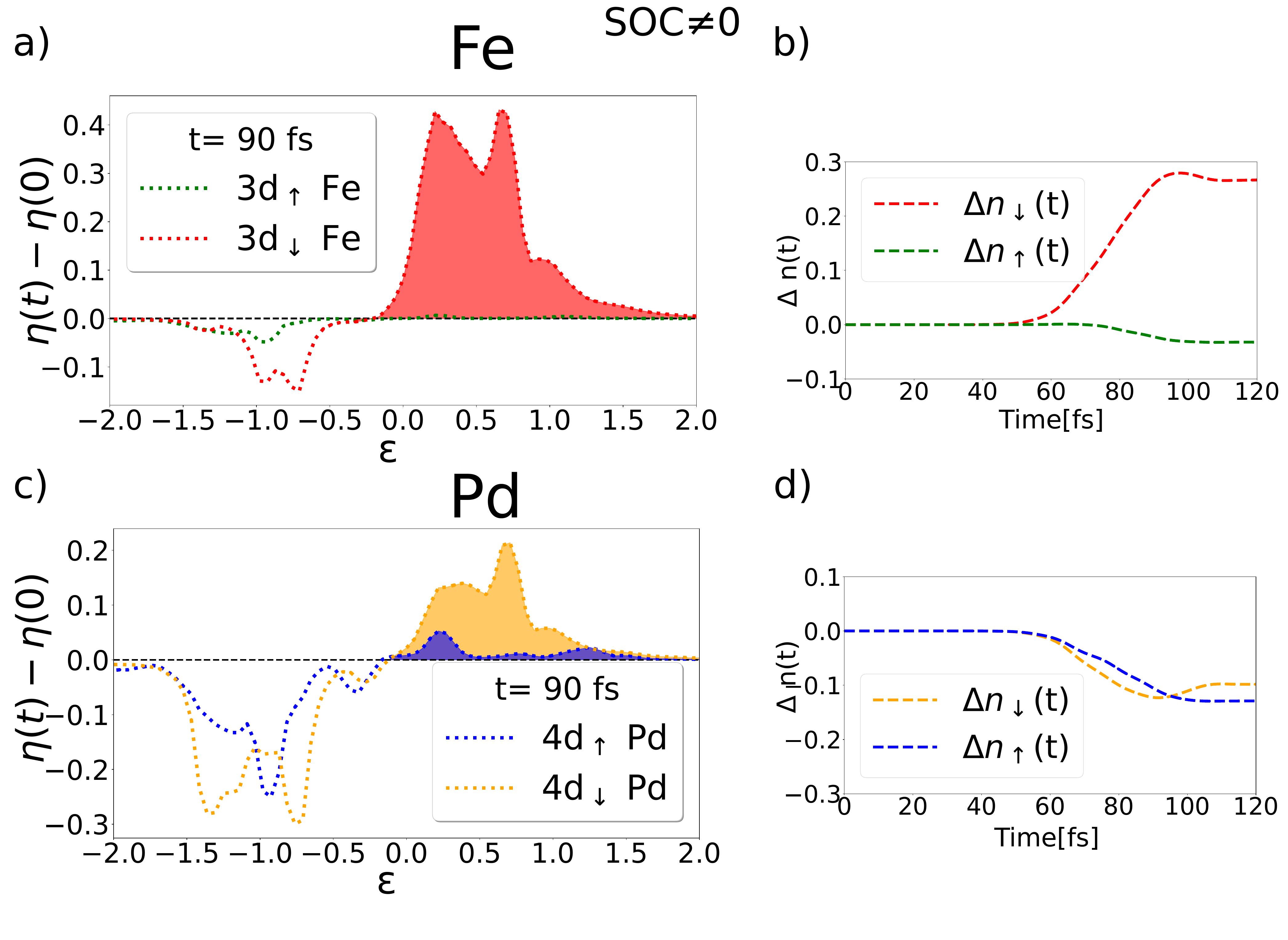}
    \caption{Calculations with finite SOC.(\textbf{a,c}) the change in the occupation of the 3d up and down spin channels to the GS ($\eta(0)$) at t=90 fs for Fe and Pd.(\textbf{b,d}) the change in the integration of the majority($\Delta n_\uparrow$(t)=$\int d\epsilon (\eta_\uparrow(\epsilon,t)-\eta_\uparrow(\epsilon,0))$) and minority($\Delta n_\downarrow$(t)) spins occupations over the entire energy range in time for Fe and Pd. }
    \label{fig:change_pdos_soc}
\end{figure}

To analyze further the microscopic origin of the OISTR process in element specific fashion, the time dependent change in the majority (spin up $\uparrow$) and minority  (spin down $\downarrow$) spin occupations of the d level, $\eta_d(t)$, is calculated using Eq.\ref{eq:td_pdos}. The  difference between  $\eta_d(t)$ at t=90 fs  and   t=0  \big ($\eta_d(0)$\big ) is depicted in Fig. \ref{fig:fepd_pdos_nosoc} (a and c) and \ref{fig:change_pdos_soc} (a and c) for SOC=0 and  SOC$\ne$0, respectively. The integration of this quantity over the entire energy range [$\Delta n(t)=\int d\epsilon (\eta(\epsilon,t)-\eta_(\epsilon,0))$] is shown as function of time, in Fig.  \ref{fig:fepd_pdos_nosoc}\textcolor{blue}{b} and \ref{fig:fepd_pdos_nosoc}\textcolor{blue}{d} for SOC =0 and in Fig.     \ref{fig:change_pdos_soc}\textcolor{blue}{b} and \ref{fig:change_pdos_soc}\textcolor{blue}{d} for SOC $\ne$0. For the SOC=0 case, the decrease in the initially occupied minority spins of Pd (unshaded yellow area of Fig. \ref{fig:fepd_pdos_nosoc}\textcolor{blue}{c}) and Fe (unshaded red area of Fig. \ref{fig:fepd_pdos_nosoc}\textcolor{blue}{a}) is transferred to the initially unoccupied  minority spin channel of Fe (shaded red area of Fig.  \ref{fig:fepd_pdos_nosoc}\textcolor{blue}{a}) and Pd (shaded yellow area of Fig.  \ref{fig:fepd_pdos_nosoc}\textcolor{blue}{a}), respectively. This can be pinpointed as a pure OISTR effect because SOC is turned off. The amount of the transferred spins is in this situation consistent with the GS empty states available in the minority spin channels for every species of Fig. \ref{fig:gs_d_orbital_dos}\textcolor{blue}{a}. Note that $\Delta n_\downarrow(t)$ in Fig. \ref{fig:fepd_pdos_nosoc}\textcolor{blue}{b} and \ref{fig:fepd_pdos_nosoc}\textcolor{blue}{d} are consistent with this observation. That is to say the OISTR process of transferring electron states from Pd to Fe and vice versa is selective only through the minority spin channel. 

For SOC$\ne$0 case, in addition to OISTR, a signature of spin flip transitions in the spin up channel for both Fe (unshaded green line in Fig. \ref{fig:change_pdos_soc}\textcolor{blue}{a}) and Pd (unshaded blue line \ref{fig:change_pdos_soc}\textcolor{blue}{c}) is manifested. The spin flips by SOC are more intense in Pd due to its stronger SOC.

A striking feature in Fig. \ref{fig:fepd_all_mom_compar}\textcolor{blue}{b} is that for calculations with finite SOC, one notices that the onset of the demagnetization of Pd (green graph) comes earlier than the onset of Fe demagnetization (black graph). While the time scale at which SOC starts to cause demagnetization of magnetic material is an intrinsic property, the time scale at which OISTR can be activated is an extrinsic property, i.e it can be controlled by the laser pulse parameters. Having identified OISTR and SOC as the two major sources for details in the demagnetisation in FePd$_3$, a relevant question is whether one can utilize the laser pulse parameters to selectively manipulate the onset of demagnetization in an element specific way. To investigate this we have considered a laser pulse with the same fluence as that considered in Fig.\ref{fig:fepd_all_mom_compar}, but with a shorter FWHM. We have also considered significantly larger fluences, up to 13 mJ/cm$^2$.
\begin{figure}
    \centering
    \includegraphics[width=\linewidth]{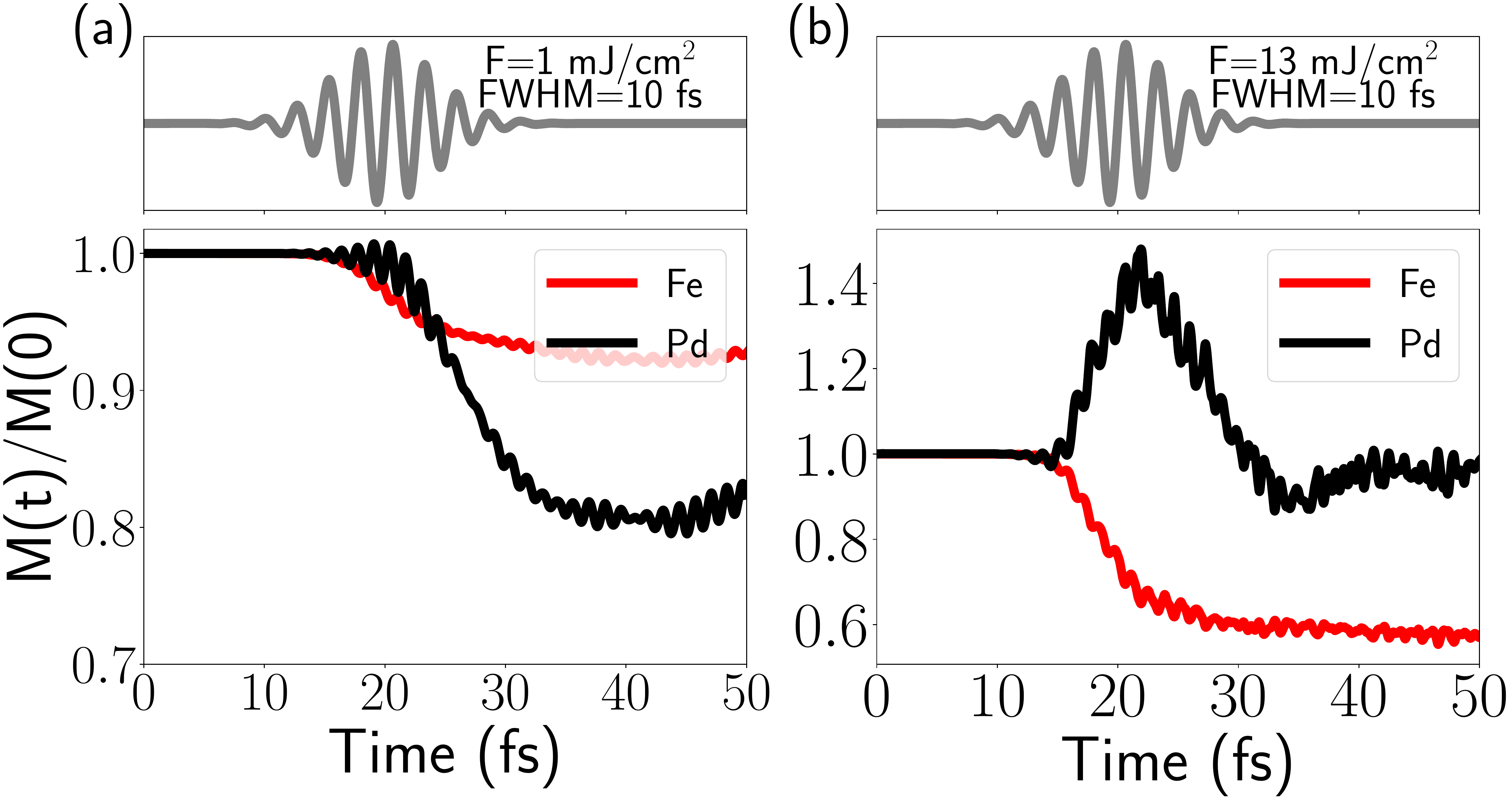}
    \caption{(\textbf{a}) Pulse shape of the 1mJ/cm$^2$ laser with FWHM=10 fs(upper panel).(\textbf{b}) pulse shape of the 13 mJ/cm$^2$ with FWHM=10 fs (upper panel). The lower panels show the moment loss in Fe and Pd under the influence of the corresponding pulse. }
    \label{fig:short_ultrahshort_mom}
\end{figure}

Figure \ref{fig:short_ultrahshort_mom}  shows results from calculations in which SOC was finite. In the figure the demagnetization of Fe and Pd is displayed, using a short pulse of FWHM=10 fs and fluence of 1 mJ/cm$^2$ (Fig.  \ref{fig:short_ultrahshort_mom}\textcolor{blue}{a}) and stronger pulse of 13  mJ/cm$^2$ with the same FWHM (Fig. \ref{fig:short_ultrahshort_mom}\textcolor{blue}{b}). In both cases, Fe is seen to have an onset of demagnetisation that comes earlier than Pd, which is in contrast to the results of figure \ref{fig:fepd_all_mom_compar}.

The trend exhibited by the three sets of calculations is rather clear. For weaker, longer pulses, in which the vector potential $\mathbf{A}_{ext}(t)$ in Eq.\ref{eq:TDSERG} is less influential Pd demagnetises before Fe, with a relative reduction of the moment that is strongest for Pd. For the case with intermediate values of $\mathbf{A}_{ext}(t)$ (Fig.  \ref{fig:short_ultrahshort_mom}\textcolor{blue}{a}), the Fe moment demagnetises before Pd, but Pd has an overall more pronounced relative demagnetization. For the strongest pulse (Fig.\ref{fig:short_ultrahshort_mom}\textcolor{blue}{b}), the moment on Pd actually increases initially, by 40 \%, before it decreases and becomes reduced with respect to its GS value. In this case Fe demagnetises more than Pd. 
One can understand this trend by noting that for calculations with very large fluence, that operate during short time, $\mathbf{A}_{ext}(t)$ seems to dominate over SOC, so that spin-flip transitions become unimportant compared to the dominating OISTR effect. For this reason, the results in Fig. \ref{fig:short_ultrahshort_mom}\textcolor{blue}{b} are similar to the data without SOC in Fig.\ref{fig:fepd_all_mom_compar}\textcolor{blue}{b}. Gradually decreasing the strength of $\mathbf{A}_{ext}(t)$ diminishes the importance of the OISTR effect, compared to spin-flip transitions, which is why in Fig.\ref{fig:short_ultrahshort_mom}\textcolor{blue}{a} the Pd moment stays constant initially, while the Fe moment decreases. In Fig.\ref{fig:fepd_all_mom_compar}\textcolor{blue}{b}, the influence of $\mathbf{A}_{ext}(t)$ is reduced significantly, such that the Pd moment demagnetises first and most.

Interestingly, the LIUMD of this hybrid Stoner-Heisenberg system is  unexpectedly dominated by the OISTR  effect in conjunction with SOC mediated spin-flip excitations for at least 140 fs. The demagnetization of Pd  was found to be mainly driven by its strong SOC, while in Fe, both SOC and OISTR  provide significant channels of demagnetization. The rigidity of the Fe-moment, resulting from the strong intra-atomic exchange coupling on Fe, inhibits local spin-flip excitations ($J_{\text{ex}}>>\text{SOC}$), and the demagnetization process is strongly influenced by the ability to transfer spin-density to the nearby Pd atom. This mechanism facilitates the ability to site-selectively engineer demagnetization by purely optical means. Site-selective control of the demagnetization process provides the fastest way to engineer the demagnetization dynamics in alloys that is desirable for spintronics technology. 

Remarkably, the drastic difference in origin of the magnetic moment of the Fe and Pd species is not a determining factor for the magnetization dynamics in the first 100 fs. Magnetization dynamics on longer time-scales are bound to be influenced by magnon excitations, and we suggest that at such time-scales the known difference in processes determining magnetic excitations should play role for the dynamics. We also suggest that the non-equilibrium magnetization density thermalizes through magnon excitations, where the local exchange splitting is allowed to relax. 








\section{Acknowledgements}
The computations were enabled by resources provided by the Swedish National Infrastructure for Computing (SNIC) at NSC and Uppmax partially funded by the Swedish Research Council through grant agreement no. 2018-05973. OG acknowledge financial support from the Strategic Research Council (SSF) grant ICA16-0037 and the Swedish Research Council (VR) grant 2019-03901. This work was also supported by the European Research Council via Synergy Grant 854843 - FASTCORR.

\bibliography{ultrafast_demag_Fe_Pd}
\bibliographystyle{apsrev4-1}

\end{document}